# Near room-temperature memory devices based on hybrid spin-crossover@SiO$_2$ nanoparticles coupled to single-layer graphene nanoelectrodes

Anastasia Holovchenko, Julien Dugay*, Mónica Giménez-Marqués, Eugenio Coronado* and Herre S. J. van der Zant

A. Holovchenko, Dr. J. Dugay, Prof. H. S. J. van der Zant

Kavli Institute of Nanoscience, Delft University of Technology, Lorentzweg 1, 2628 CJ Delft, the Netherlands

E-mail: j.dugay@tudelft.nl; h.s.j.vanderzant@tudelft.nl

Dr. M. Giménez-Marqués, Prof. E. Coronado

Instituto de Ciencia Molecular, Unversidad de Valencia, Catedrático José Beltrán 2, 46980 Paterna, Spain

E-mail: eugenio.coronado@uv.es

In weak crystal fields octahedral complexes based on $3d^4$–$3d^7$ transition-metal ions often display a spin-crossover (SCO) between their high-spin (HS) and low-spin (LS) electronic configuration that can be induced by various external stimuli such as light, temperature, pressure, guest molecules, magnetic field and an electric field.[1,2] In addition, in the solid state some of these systems exhibit a remarkable memory effect as a result of strong cooperative elastic interactions occurring between metal centers.[3]

In the last five years, several research groups have evidenced large memory effects in SCO devices by measuring their transport properties either in powdered samples[4–6] or in micro and nanostructures [7,8][9][10] going down to the single nanoparticle (NP) level.[11] Notably, electrical control over the SCO has given an enormous impulse to this field,[11] and SCO NPs, first reported by some of us,[12] are nowadays considered as promising candidates to be used as active parts in molecular-based memory devices.

A practical use of these SCO devices, however, has so far been hampered by the low reproducibility of the hysteresis loop in the conductance. This is particularly dramatic when one or a few NPs are contacted between electrodes for room-temperature (RT) operation.[7,11] In fact, these devices rapidly degrade above RT and the thermal hysteresis loop often disappears after the first electrical-thermal cycle. To the best of our knowledge, the only example published of reproducible memory effect in the conductance at the nanoscale was demonstrated through 4 switches in the current-voltage characteristics operated at low temperatures (10 K) using [Fe(Htrz)$_2$(trz)(H$_2$O)](BF$_4$) NPs coated with an organic surfactant.[11] Using interdigitated electrodes, Lefter et al. [9] measured up to 20 thermal hysteresis loops of the electrical current above RT for larger assemblies of organized micro-rods of the same SCO compound. However, even in these large assemblies a progressive degradation was systematically observed upon each thermal cycle (i.e continuous current decrease). The loss of particle/particle or particle/electrode contacts may be the main cause of this electrical degradation, although one can imagine that the genuine fatigue of these compounds could be influenced in many other ways (temperature rate, voltage, intrinsic fatigue, etc.). A systematic diversification of both the probing techniques and physical conditions, as well as the nature and/or the range of the external stimuli applied should be considered to unveil and individually identify the mechanisms of fatigue.

In this context, graphene[13] may serve as an interesting new material for the electrodes. With only one atomic layer thin, graphene electrodes are non-invasive, and conduct both heat and electricity very efficiently, while being at the same time thermally stable even above RT.[14] In addition, graphene is optically transparent and can efficiently guide surface plasmon modes that can be dynamically tuned by electrostatic gating.[15] The latter feature, in synergy with the memory effect in the dielectric properties of triazole-based SCO compounds[6] holds great promise to manipulate active plasmonic devices by an appropriate external perturbation.[16] All these characteristics make graphene a promising material for use as electrode in devices based on SCO NPs.

The coating layer, usually stabilizing sub-micrometric SCO NPs can also play a major role in their chemical stability, and thus in retaining memory-effect features. In particular, Colacio et al. [17] proposed hybrid SCO NPs using silica ($SiO_2$) as a robust inorganic shell. These hybrid core-shell NPs already proved to be multi-functional when decorated with luminescent molecules[17] or gold NPs displaying both electronic bistability and luminescence[18] or plasmonic properties, respectively.[17,19]

Recently, in a different approach the flexibility of conducting polymers (polypyrrole) has been reported to circumvent the problem of reproducibility. Galán-Mascarós et al. [20] prepared 50 μm-thick composite films made of similar triazole-based SCO compounds embedded in such conducting polymers. The conduction level of the latter turned out to be very sensitive to the thermally induced volume change of the SCO compounds, leading to a conduction change in a range of 50-300%. However, this alternative strategy will become increasingly less suitable as device size decreases down to the nanometer regime, as strongly grafted conducting polymers on active SCO cores is not yet achieved.

Here, we report on the charge transport properties of SCO [Fe(Htrz)$_2$(trz)(H$_2$O)](BF$_4$) NPs covered with a silica shell placed in between single-layer graphene electrodes. We evidence a reproducible thermal hysteresis loop in the conductance above room-temperature. This bistability combined with the versatility of graphene represents a promising scenario for a variety of technological applications but also for future sophisticated fundamental studies.

**RESULTS**

Single-layer graphene electrodes were defined with electron-beam lithography and oxygen plasma etching on commercially available Si/SiO$_2$/CVD-graphene substrates (see Figure 1(a) and Experimental Section for more details). Several combinations of the electrode-separation width to length ratios (W/L as defined in Figure 1(b)) were fabricated, enabling to contact a single NP with L ≈ W ≈ the particle size, up to small assemblies of hybrid SCO NPs (representing a couple of NPs in series and about 30 possible parallel electrical pathways at maximum). A Scanning Electron Microscope (SEM) image of representative device is shown in Figure 1(b).

Core-shell SCO NPs of *ca.* 110 nm in length and *ca.* 50 nm in width were synthesized using published methods[12] (see Experimental Section for more details). The SCO core was obtained from the [Fe(Htrz)$_2$(trz)(H$_2$O)](BF$_4$) (Htrz = 1,2,4-triazole and trz = 1,2,4-triazolato) complex and a shell of SiO$_2$ was grown around it. The thermal spin-transition of the powder sample was magnetically established using a SQUID magnetometer (Figure SI1). A suspension of the same powder was prepared in ethanol and drop-casted on top of the devices. A dielectrophoresis technique was used to trap the NPs in between the graphene electrodes (see Experimental Section for more details). SEM micrographs taken after deposition at an acceleration voltage of 10 kV show that in 85% of the devices (42 devices out of 49) few or even a single NP(s) was (were) placed between the two electrodes (see representative examples in Figure SI5).

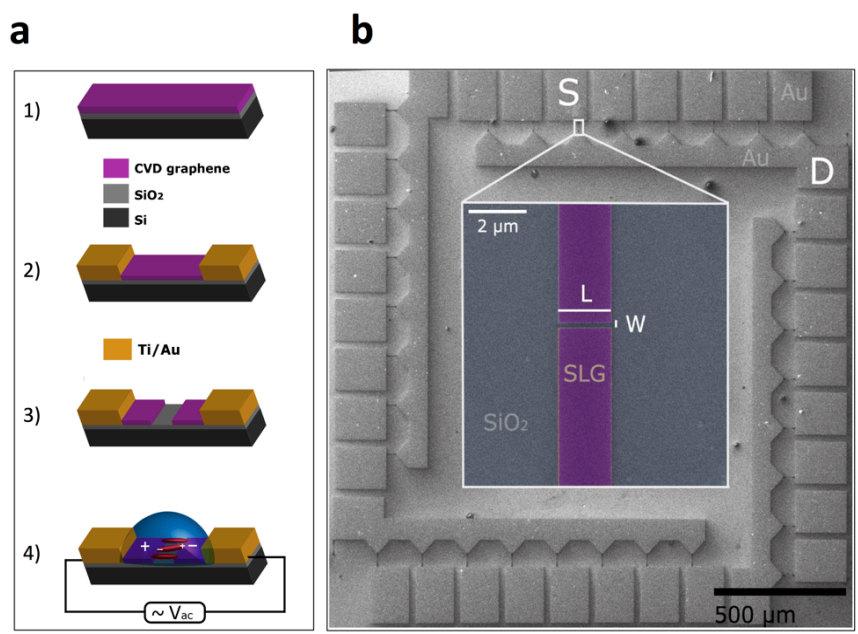

**FIGURE 1** – **(a)** Schematic of the fabrication process flow. Step 1-2: Gold contact pad and lead definition by electron-beam lithography, oxygen plasma etching and metal evaporation (see Experimental Section for more details). Step 3: Nanogap formation by electron-beam lithography and oxygen plasma etching. Step 4: Electrical trapping of the hybrid spin-crossover@SiO$_2$ NPs by a dielectrophoresis technique. **(b)** Scanning-electron microscopy micrograph of a chip before NP deposition containing 32 devices. **Inset:** Single-layer graphene electrodes of length L = 1.7 μm and separation width W = 300 nm on a Si/SiO$_2$ substrate (false colour).

Electrical characterization of the devices was first performed before deposition. Current-voltage characteristics (I-Vs) were taken at atmospheric pressure using a cryostat probe station (Desert Cryogenics). All recorded I-V curves showed a maximum leakage current of about 5 pA while sweeping the bias voltage up to 35 V. Temperature-dependent conductance measurements before deposition showed no conductance change as a function of temperature between 305 K and 375 K (Figure SI6). For these measurements a heater element was embedded in the probe station with a Lakeshore temperature controller and a local calibrated thermistor (TE-tech, MP-3011). Heating was performed at a rate of 2 K·min$^{-1}$ by means of a resistor underneath the sample stage. Cooling was not done actively, but the rate could be controlled at high temperatures (from 385 K to 340 K), where the natural thermal dissipation exceeded 2 K·min$^{-1}$.[21]

I-Vs recorded after deposition (same conditions as mentioned above) showed an increase of the current in 23 % of the devices (8 out of 34). For these devices temperature-dependent conductance measurements were performed to investigate the spin-state dependence of the charge transport properties. It should be noticed that an increase of current was only seen in the devices with a large W/L ratios. We observed no change in current after deposition for the devices, which had W/L equal to the size of the single NP, which could indicate that multiple parallel pathways are needed to observe a distinguishable current increase for these NP systems.

From SQUID measurements performed on a powder sample (see Figure 1 SI) it is known that switching between the two spin states is expected in the temperature range from $T_{1/2}^{\uparrow}$ = 380 K to $T_{1/2}^{\downarrow}$ = 340 K ($T_{1/2}^{\uparrow}$ and $T_{1/2}^{\downarrow}$ = temperature for which there are 50% high-spin and 50% low-spin states in the heating and cooling mode, respectively). Figure 2(a) displays a thermal hysteresis loop in the conductance of the sample A (W = 150 nm, L = 1.7 μm) at an applied bias voltage V = 20 V. The hysteretic behaviour of the conductance as a function of temperature is consistent with the temperatures observed in the powder sample.

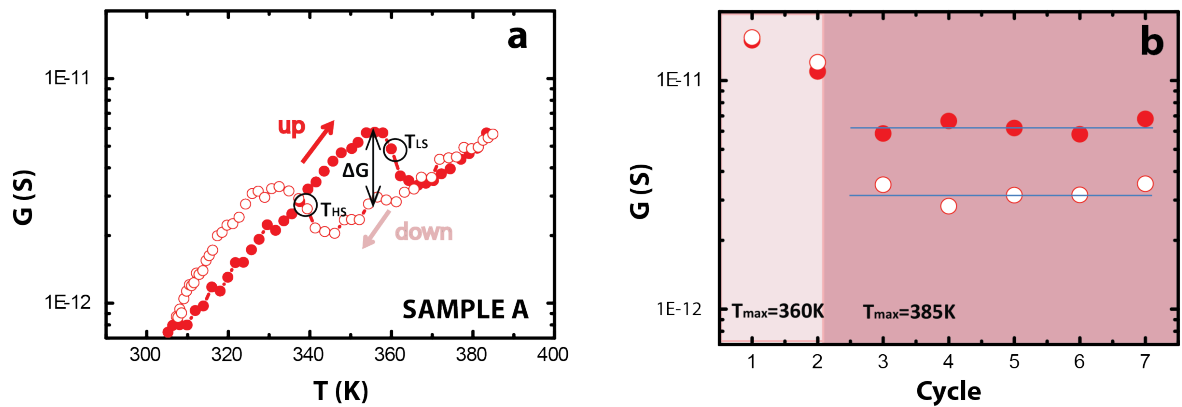

**FIGURE 2** – **(a)** Temperature-dependent conductance (G = I/V) measurements of sample A (cycle 6) showing hysteresis between the heating up (solid red circles) and cooling down (open white circles) cycle. **(b)** Conductance of sample A at T = 355 K and V = 20 V in the heating (solid red circles) and cooling (open white circles) modes for several cycles. Cycle 1-2: Maximum temperature while sweeping (heating mode) $T_{max}$ = 360 K is too low for the spin switching to occur. Cycle 3-7: $T_{max}$ = 385 K.

Up to seven cycles (of sample A) were performed to check reproducibility of hysteresis features (Figure 2(b)). During the first two cycles the maximum temperature ($T_{max}$) was set to 360 K. Under these circumstances no conductance change between the heating and cooling modes was observed, indicating that 360 K is a too low temperature for spin switching to occur. When $T_{max}$ was increased up to 385 K, a well-pronounced and reproducible hysteresis feature was observed (cycles 3-7 in Figure 2(b)). The conductance in the heating mode appears to be higher than in the cooling mode (open white and solid red circles, respectively). Importantly, for all five cycles the conductance values remain stable as well as the critical temperatures, which show that the spin switching does not degrade upon temperature cycling. Figure 3(a) and (b) show the presence of hysteresis loops for sample B and sample C, respectively. These samples have a slightly different L/W ratio of the graphene electrodes, but the hysteresis behaviour remains a common feature of the data. Indeed, the critical temperatures for both the heating and cooling modes are very close for samples A, B and C and also in good agreement with the transition temperatures obtained on the powder sample (see Figure SI1 for more information). Remarkably, the width of the hysteresis loop ΔT and the relative change in conductance ΔG are increasing with the separation width (W) of the graphene electrodes. Table 1 summarizes the parameters extracted from the hysteresis loops of the three devices (for sample A see Figure 2(a), for sample B see Figure 3(a) and sample C see Figure 3(b)). It is important to note, that the low-conducting state is the HS state for all three samples.

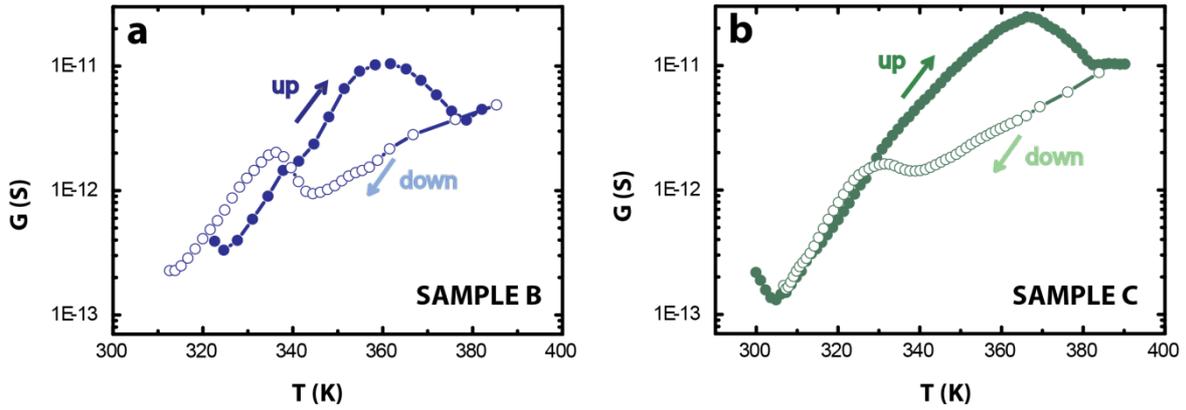

**FIGURE 3** – Conductance (G = I/V) as a function of temperature for **(a)** sample B at V = 30 V and **(b)** sample C at V = 35 V. Hysteretic behaviour of the conductance indicates NP switching between the low-spin state and high-spin state.

| Sample | L(μm) | W(nm) | ΔG (nS) | $T^{LS}$ (K) | $T^{HS}$ (K) | ΔT(K) |
|---|---|---|---|---|---|---|
| A | 1.7 | 150 | 2.18 | 340 | 360 | 20 |
| B | 0.7 | 250 | 5.88 | 340 | 374 | 33 |
| C | 1.7 | 300 | 6.11 | 335 | 374 | 39 |

**TABLE 1** – Sample dimensions and parameters deduced from the conductance versus temperature plots. W and L are respectively the separation width and length of the graphene electrodes, ΔT is width of the hysteresis loop and ΔG is the difference in conductance between the high- and low-spin state. For the definition of the other parameters, see the main text and Figure 2 (a).

In the literature, conductance vs. temperature curves have been fitted to exponential Arrhenius dependencies. We extracted activation energy values ($E_a$) and pre-exponential factors ($G_0$) in both spin-states from an Arrhenius fit through the data (see also Fig. 4):

$$\ln G = \ln G_0 - \frac{E_a}{k_B T}. \qquad (1)$$

Note, that we have used the conductance and not the conductivity in this equation, as we do not exactly know the current pathways in our samples.

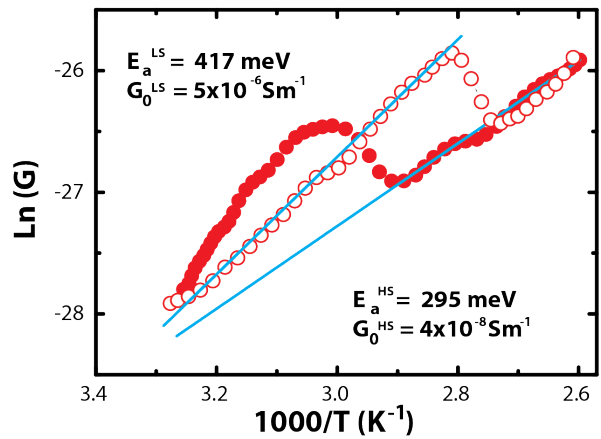

**FIGURE 4** − Arrhenius plot of the logarithm of the conductance vs. the inverse temperature of sample A. The activation energy is obtained by fitting a linear curve through the high-spin data (solid red circles) and the low-spin data (open white circles).

The parameters obtained from the Arrhenius fits for sample A, B, C are given in Table 2. For all three samples $E_a$ and $G_0$ are both higher in the low-spin state compared to the high-spin state. Besides, the variation in the $G_0$ values is much larger (several orders of magnitude) than that in the $E_a$ values.

| Sample | $E_a^{LS}$ (meV) | $E_a^{HS}$ (meV) | $G_0^{LS}$ (S m$^{-1}$) | $G_0^{HS}$ (S m$^{-1}$) |
|---|---|---|---|---|
| A | 417 | 295 | $5 \times 10^{-6}$ | $4 \times 10^{-8}$ |
| B | 1085 | 501 | $2 \times 10^{4}$ | $2 \times 10^{-5}$ |
| C | 887 | 464 | 56 | $1 \times 10^{-5}$ |

**TABLE 2** – Thermal activation energies ($E_a$) and pre-exponential factors ($G_0$) of the conductance in the two spin-states (HS – high spin state, LS – low spin state) for sample A, B and C. For parameter definitions see main text and Figure 3(a).

**DISCUSSION**

In this paper we demonstrated, for the first time, that graphene electrodes can be employed to probe phase-transitions occurring near RT in SCO compounds. A dielectrophoresis deposition method[7] turned out to be adequate method to position the small hybrid SCO@SiO$_2$ NPs in between the single-layer graphene electrodes. Importantly, we have observed reproducible hysteresis in the conductance upon thermal cycling for 5 times above RT. No degradation of the current levels has been detected; we thus conclude that the robustness of the spin-transition is significantly improved as compared with our previous results[11,21] and with other reports for larger assembles and SCO objects.[9]

As possible reasons influencing the spin-transition instability one can envision the following causes: *i)* a progressive breaking of the molecular structure induced by thermal cycling, as shown recently by Grosjean et al.,[22] albeit not demonstrated at the nanoscale, *ii)* a modification in the charge of the metal centres/counter ions induced by the flow of charge carriers and/or sample environment,[6,23] *iii)* frictional forces at the NP/substrate interface[21,24] or *iv)* a loss of electrical contact upon temperature cycles.[9] In distinguishing between these different effects, the silica matrix could be beneficial as it decouples the active cores from the substrate and environment. Further studies are thus needed and in particular one should reproduce the same studies while scaling down the sizes of the hybrid NPs. Moreover, the atomically thin graphene electrodes could yield improved electrical contact with the hybrid NPs upon thermal cycles.[14]

Another important observation is the spread of the hysteresis loop widths, which seems at a first glance in contradiction with recent results obtained on powder samples. Indeed, our loop width value has increased to 19 K using the same synthesis batch for lengths and widths respectively ranging from 109 $\pm$ 24 nm and 45 $\pm$ 30 nm; in contrast, Colacio et *al.*,[17] reported hysteresis loop widths of 5 K only while comparing four different syntheses with NPs having lengths and widths respectively ranging from 56-422 nm and 56-180 nm. We ascribe this notable difference to the fact that we electrically probed only few NPs, whereas they used a SQUID magnetometer to magnetically characterise a large amount of material. Consequently, the changes in cooperativity due to different NP sizes, morphologies and silica thickness could be masked due to averaging.

As far as the transport mechanism in the SCO NPs is concerned, the Arrhenius fit to the data indicates a broad variation in the thermal activation energies and pre-exponential factors. A similar large spread was reported in the literature[25] associated to the different morphologies adopted by the same SCO compound, depending on the synthesis procedure. In contrast, the single-batch synthesis employed here points out that a large spread of the pre-exponential factor and activation energies is already present and linked to the inherent core size distribution, the silica shell thickness variation and/or local structural and/or chemical defects.[4]

More importantly, we systematically found that $E_a^{LS} > E_a^{HS}$ and $G_0^{LS} > G_0^{HS}$ while the opposite trend was reported for micro-rods made of the same SCO compounds that are not stabilized by a shell.[4,26] On the other hand, the low-conducting HS state reported in our work remains, however, consistent with previous studies at the macro- and nanoscale.[4,6–10,21,26] In this context, we speculate that the inversion in thermal activation energies and pre-exponential factors can be rationalized by the presence and the compression/relaxation of the silica shell accordingly to the expansion/relaxation of the NP cores upon SCO. Specifically, $E_a$ and $G_0$ may originate from a contribution of hopping through both the core and the shell:

$$\ln G = \ln G^{core} + \ln G^{shell}, \quad (2)$$

where $\ln G^{core}$, $\ln G^{shell}$ are of the same form as equation.1, $E_a = E_a^{core} + E_a^{shell}$ and $G_0 = G_0^{core} \times G_0^{shell}$. Regarding the core contribution, it has already been evoked that its pre-exponential factor $G_0^{core}$ is governed by the competition of the product of the hopping distance and the hopping frequency.[4] Within the NP cores in the HS state, the well-known metal-ligand bond increases and softening will favour a charge carrier to hop further through the core, but with less probability. Similarly, we propose that the pre-exponential factor of the shell $G_0^{shell}$ could be governed by the same competition, but yielding the opposite behaviour: in the HS state, charge carriers hop less far upon the silica compression but with higher probability.

Thermal hopping activation energies are even more complicated to capture,[4] as it refers to the reorganization processes of atoms and molecules occurring for each charge transfer on SCO complexes. So far, higher activation energies have been reported consistently in the HS state for vast aggregates of nanometric particles[4] and micro-rods[26] made of the same SCO compound used in this work. If higher activation energies of the NP cores in the HS state hold in our case as well, it means that the activation energy for the shells has to be smaller to explain the systematic conductance drop in our measurements. This can be rationalized by the compression of the atom distances within the shell in the HS state, which is expected to decrease the activation energy. These aforementioned intricate effects can thus explain an inversion of the activation energies and pre-exponential parameters compared to previous studies and points at the deformable ability of the silica shell wrapping the active cores.

One can wonder nevertheless, whether a tunnelling contribution within the shells can be expected for hybrid SCO@SiO$_2$ NPs. Tunnelling transport will occur if the barrier height is high and thin. High-angle annular dark field scanning transmission electron microscopy (HAADF−STEM) images using tomographic acquisition of similar hybrid NPs recently revealed that the wrapping mean silica is 12 nm thick (*i.e.,* the interparticle distances could be as large as 24 nm).[17] Since no definite evidence is available yet for the hybrid NPs employed here, the coexistence of tunneling and hopping conduction paths cannot be excluded between hybrid NPs.

**Conclusion**

In this work we have successfully trapped small assembles of hybrid SCO@SiO$_2$ NPs by a dielectrophoresis technique in between single-layer graphene nanogaps. This class of SCO NPs offers a large memory effect in the conductance of about 40 K associated with the spin-transition. The hysteresis occurs concomitantly with a reproducible and efficient back-relaxation to the low-spin state as a possible result of the enhanced stability provided by the silica shell on the SCO core NP. Moreover, the low-spin state possesses a higher conductance, consistent with previous studies on small assemblies of SCO-based NPs based on the same compound [Fe(Htrz)$_2$(trz)(H$_2$O)](BF$_4$). Importantly, the analysis of the hysteresis loop features points out that the presence of the silica shell can inverse the activation energies and pre-exponential factors while keeping the LS state as the high conducting state. We ascribe this feature to the compression (relaxation) of the shell in the HS (LS) state.

The versatility of graphene combined with the thermal hysteresis loop stability of the core-shell SCO NPs opens up future experiments, such as the study of polarized spin currents, using the enhanced gate coupling expected to be provided by the use of graphene to address the NP spin-state above RT or to manipulate the spin state by electromagnetic waves at the nanoscale.

**Experimental Section**

**Graphene electrode fabrication**

Graphene-based nanoelectrodes were fabricated on Si/SiO$_2$ substrates covered with commercially available CVD-grown single-layer graphene (obtained from Graphene Supermarket). First, gold leads were defined using e-beam lithography (EBPG5000Plus, PMMA resist, dose 900 μC / cm$^2$), followed by metal evaporation (Leybold L560 evaporator) of Ti (5 nm) / Au (60 nm) and resist lift-off. Then, the surface was covered with PMMA resist and patterned with e-beam lithography to form an etch mask defining the graphene nanogaps with widths (W) ranging from 150-300 nm and lengths (L) in the 0.6-2 micron range (extracted from Scanning-Electron Microscope (SEM) images performed at an acceleration voltage of 15 kV). The unprotected areas were then etched away with oxygen plasma etching (Leybold Fluor etcher, 25 sccm, 20 W, 500 μbar) followed by resist lift-off.

**Synthesis protocol**

Hybrid spin-crossover@SiO$_2$ NPs based on the system [Fe(Htrz)$_2$(trz)(H$_2$O)](BF$_4$), where Htrz = 1,2,4-triazole and trz = 1,2,4-triazolato, were prepared with the reverse micelle approach following a previously reported method.[13] Two separate microemulsions containing the metal and ligand coordination polymer precursors with the silica precursor were first prepared. In the first microemulsion, an aqueous solution of Fe(BF$_4$)$_2$·6H$_2$O (337 mg, 1 mmol in 0.5 mL) and 0.1 mL of TEOS were added to a solution containing the surfactant Triton X-100 (1.8 mL), *n*-hexanol (1.8 mL) and cyclohexane (7.5 mL). A similar procedure was applied for the second microemulsion, comprising an aqueous solution of 1,2,4-1H-triazole (HTrz) (210 mg, 3 mmol in 0.5 mL). Both microemulsions were combined in air and the mixture was stirred to allow micellar exchange for 6 h. Finally, destabilization of the micelles upon addition of acetone promoted the precipitation of the NPs, which were then collected by centrifugation at 12000 rpm, and washed several times with aliquots of EtOH (x3). TEM images were obtained from a JEOL JEM 1010 microscope (100 kV). Sample preparation consisted on placing a drop of the NPs suspended in a solvent on a carbon coated copper grid.

**Deposition protocol**

A colloidal solution of the hybrid SCO@SiO$_2$ particles was obtained by diluting 5 mg of a powder in 5 ml of pure ethanol. This solution was diluted one hundred times. NPs were trapped in between the graphene electrodes using a dielectrophoresis method with the following parameters: 4 V peak-to-peak sine wave between source and drain at 10 kHz. SEM images were taken using a FEI Nova NanoSEM 450 microscope operating at 10 kV.

**Acknowledgements**

We thank the EU (Advanced ERC grants SPINMOL and Mols@Mols) and NWO/OCW, the Spanish MINECO (grant MAT2014-56143-R and Excellence Unit Maria de Maeztu MDM-2015-0538), the Generalidad Valenciana (PROMETEO and ISIC Programs of excellence) for financial support of this work. M.G.-M. thanks the EU for a Marie Sklodowska-Curie postdoctoral fellowship (H2020-301 MSCA-IF-EF-658224). We also thank Aurelian Rotaru for fruitful discussions.

# Supplementary information figures

**Figure SI1:** Magnetic properties of powder samples of the SCO@SiO$_2$ NPs.

**Figure SI2:** Transmission-electron microscopy images of the SCO@SiO$_2$ NPs.

**Figure SI3:** Distribution histograms of the SCO@SiO$_2$ NPs size (length and width) from Transmission-electron microscopy images.

**Figure SI4:** Distribution histograms of the SCO@SiO$_2$ NPs size from Dynamic light scattering.

**Figure SI5:** Scanning-electron microscopy images of the electrical devices after NP deposition.

**Figure SI6:** Thermal conductance of a reference electrical device before NP deposition.

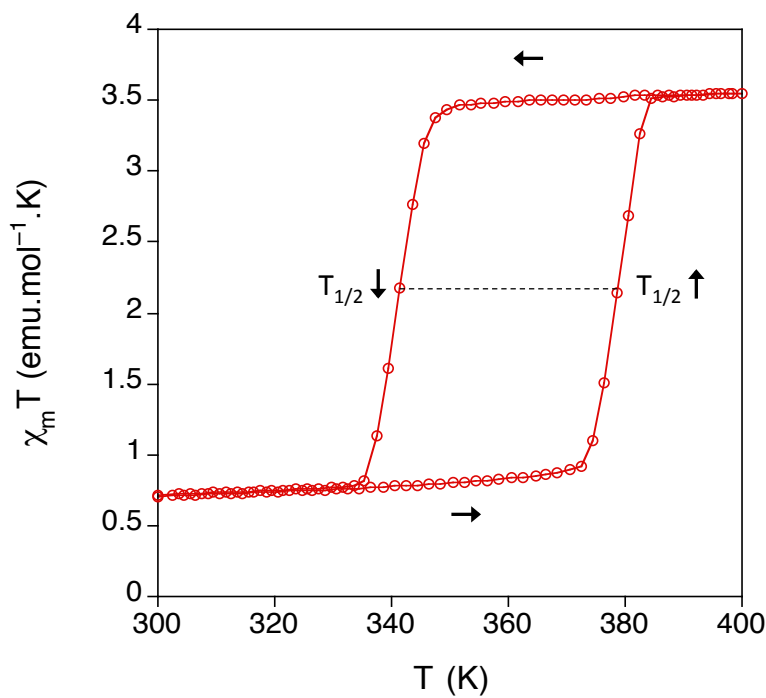

**FIGURE SI 1** –Temperature dependence of the molar magnetic susceptibility temperature product ($\chi_M \cdot T$) for powder samples of the hybrid spin-crossover [Fe(Htrz)$_2$(trz)(H$_2$O)](BF$_4$)@SiO$_2$ NPs (powder) after several heating–cooling modes.

| Powder sample | <Size> (nm) | $T_{1/2}\uparrow$ | $T_{1/2}\downarrow$ | $\Delta T$ | % HS |
|---|---|---|---|---|---|
| [Fe(Htrz)$_2$(trz)](BF$_4$)@SiO$_2$ | 107 | 379 | 340 | 39 | 20 |

**TABLE SI 1** – Physical characteristics of the thermal spin transition $T_{1/2}\uparrow$ and $T_{1/2}\downarrow$ and particle size ($T_{1/2}\uparrow$ = temperature for which there are 50% high-spin and 50% low-spin states in the heating mode).

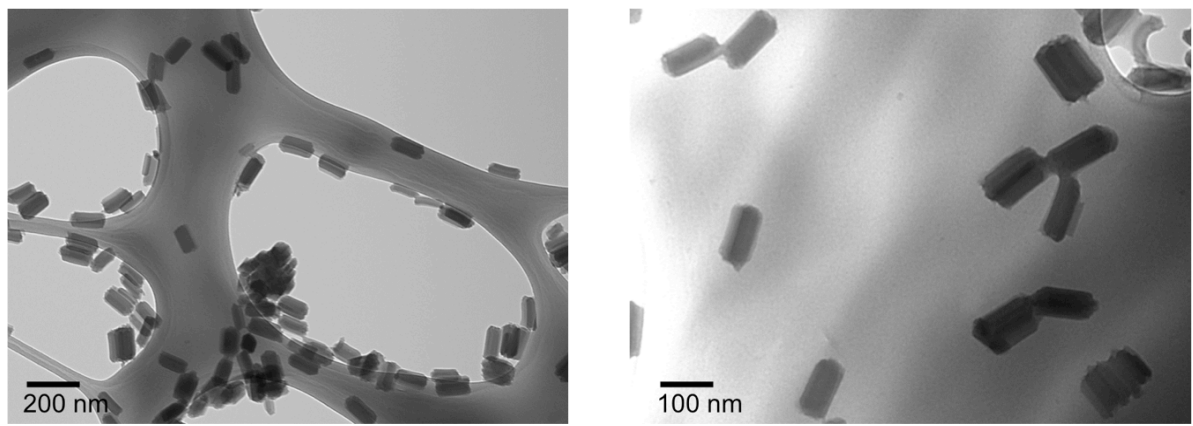

**FIGURE SI 2** – Transmission-electron microscopy images of the hybrid SCO NPs deposited by drop casting on a carbon coated copper grid.

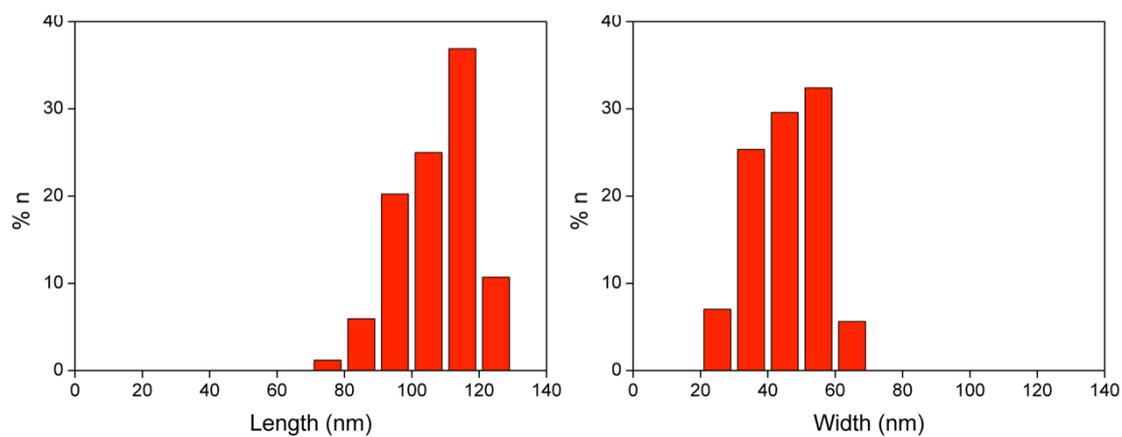

**FIGURE SI 3** − Size [length (left) and width (right)] distribution histogram and statistics for NPs of [Fe(Htrz)$_2$(trz)(H$_2$O)](BF$_4$)@SiO$_2$ obtained from TEM images (80 counts):

< length > = 107.3 nm, σ = 12.5 nm; < width > = 44.2 nm, σ = 10.1 nm

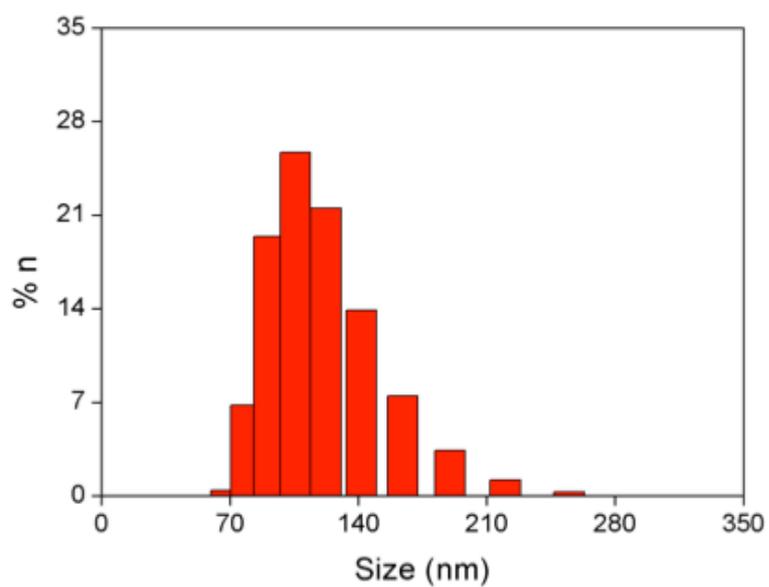

**FIGURE SI 4** – NP size distribution of [Fe(Htrz)$_2$(trz)(H$_2$O)](BF$_4$)@SiO$_2$ centered at 105 nm as determined by Dynamic Light Scattering using a Zetasizer ZS (Malvern Instruments, UK).

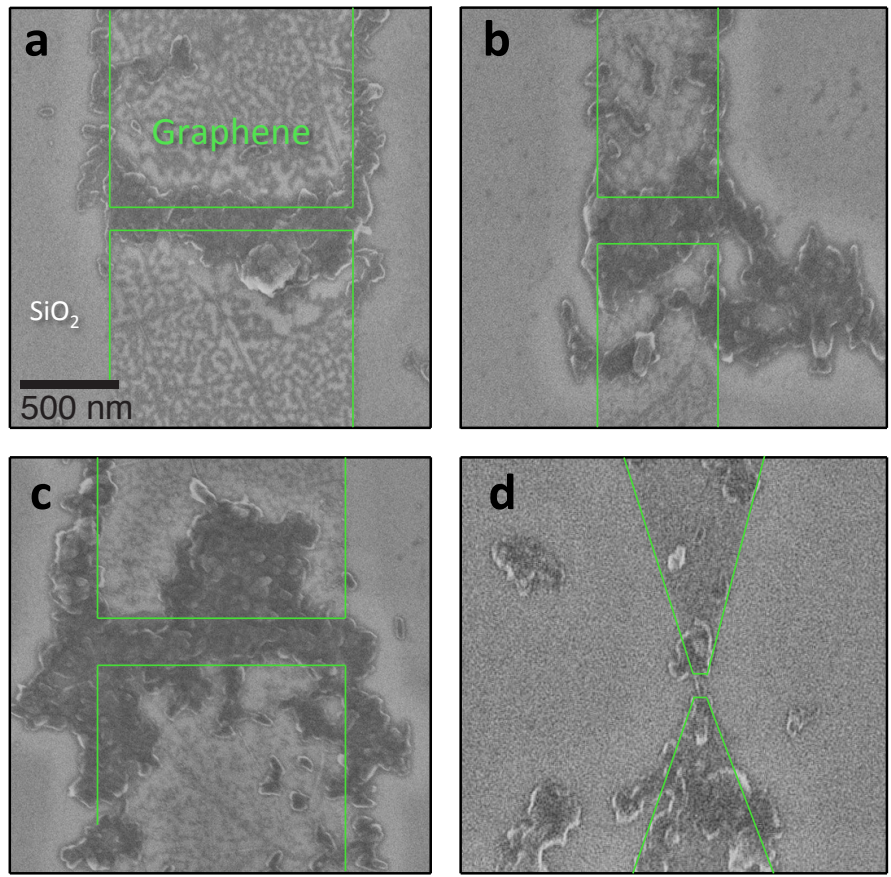

**FIGURE SI 5** – Scanning electron microscopy image of the devices after dielectrophoresis trapping. NPs were trapped in between graphene electrodes for (a) sample A (b) sample B (c) sample C and (d) sample with triangular geometry (W = 100 nm, L = 150 nm). Green lines indicate the position of the nanogap.

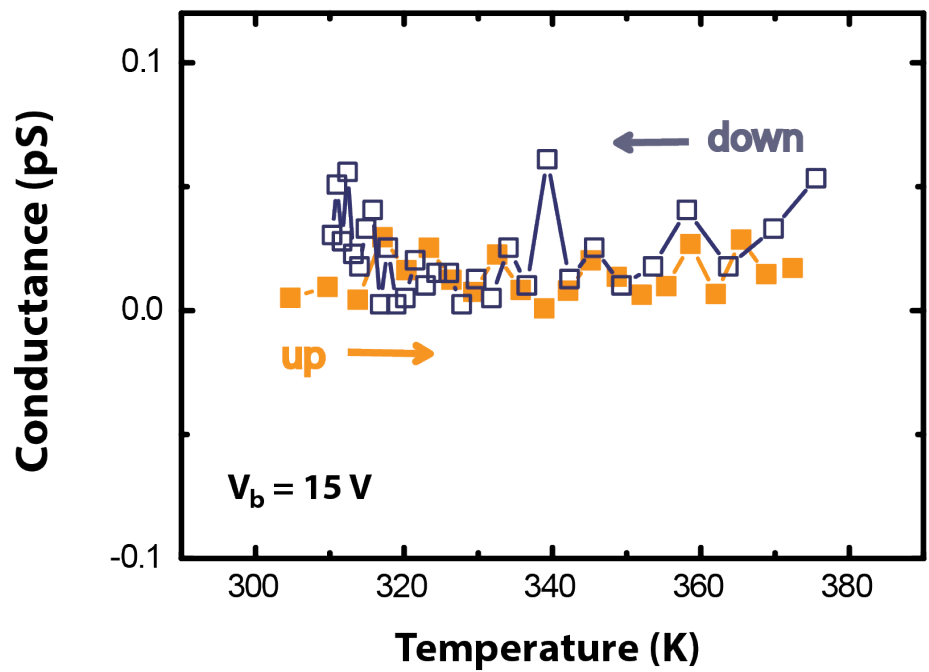

**FIGURE SI 6** – Conductance as a function of temperature for a device before NP deposition (heating mode: orange squares; cooling mode: blue empty squares). Noteworthily, the conductance levels are orders of magnitude lower than the ones in the main text after NP deposition in particular around the hysteresis temperatures.

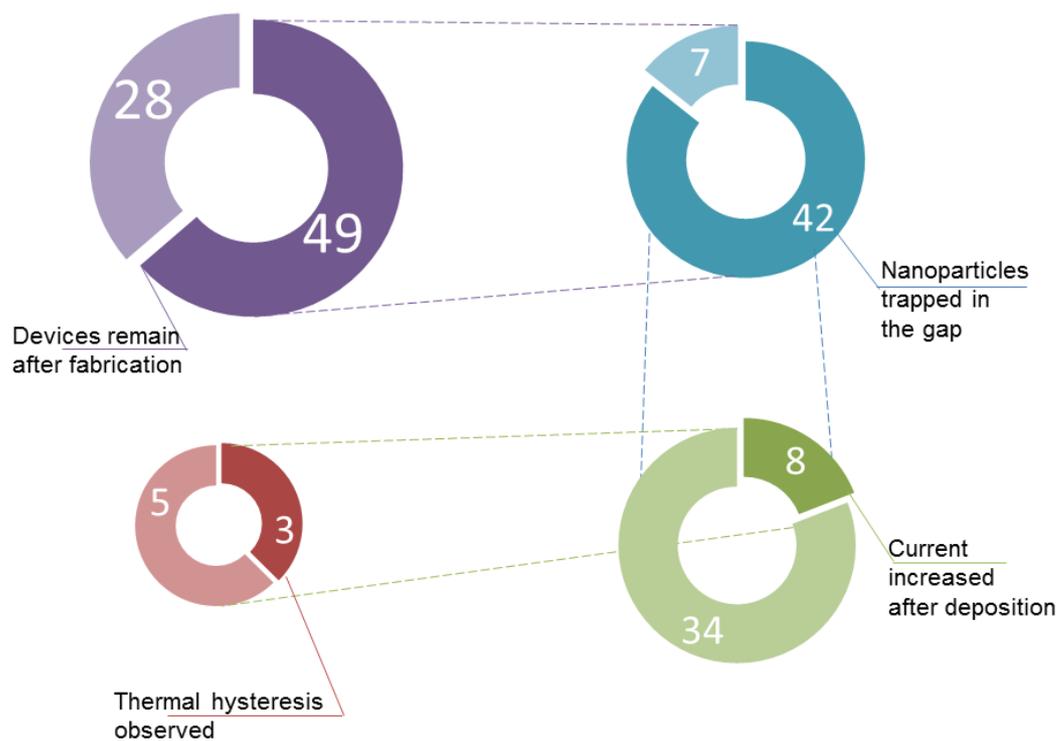

**FIGURE SI 7** – Statistics of: working/not-working devices remain after fabrication (purple chart), NP(s) trapped/not-trapped in the gaps (blue chart), electrical measurements after NP deposition (green chart) and thermal hysteresis observed in conductance (red chart).